\documentclass[11pt]{article}
\usepackage[margin=1in]{geometry}
\usepackage[utf8]{inputenc}
\usepackage[T1]{fontenc}
\usepackage{amsmath}
\usepackage{amssymb}
\usepackage{booktabs}
\usepackage{longtable}
\usepackage{array}
\usepackage{calc}
\usepackage{xcolor}
\usepackage{xurl}

\usepackage{hyperref}
\providecommand{\tightlist}{%
  \setlength{\itemsep}{0pt}\setlength{\parskip}{0pt}}

\title{What Does an RF Fingerprint Identify? A Cross-Corpus Evaluation of Hardware-Family-Associated and Unit-Level Signal}
\author{Hyojung Han \\ \texttt{hyojunguy@gmail.com}}
\date{2026-09-17}

\begin{document}
\maketitle

\begin{abstract}
RF fingerprinting benchmarks report how well a classifier separates transmitters, not whether that
separability reflects the individual unit or its hardware family. We compare a benchmark's real
hardware grouping with size-matched random groupings of the same devices, using calibrated,
unclipped variational lower bounds under label-blind, device-level cross-fitting on devices never
seen in training. On WiSig (OFDM Wi-Fi), the generation grouping exceeded all 999 matched random
groupings (permutation p = 0.001, the resolution floor), although no grouping attained a positive
certified bound, and the direction survives removing channel equalization. A preregistered
replication on a LoRa corpus (chirp spread spectrum, three hardware models), with a front-end frozen
on synthetic data, again exceeded all 999 (p = 0.001), and two sensitivity analyses judged by a
reversal rule fixed in advance did not reverse the direction. This is a directional replication
only: the excess magnitude reflects a miscalibrated null, the chain-rule decomposition leaves a
residual of +0.296 bits, and capture order is undocumented. In a same-model corpus of 122 ESP32
radios, a temperature-calibrated held-out bound of +3.338 bits (label entropy 6.931 bits) indicates
substantial device-specific identity, not separable from per-device capture-session effects; relying
on the calibrated bound was decided after the uncalibrated bound changed sign. On synthetic
transmitters a high-capacity convolutional decoder reaches 98.5\% of a known 6-bit ceiling, and the
audit protocol leaves chain-rule residuals of +0.072 bits (synthetic, fixed physical features) and
+0.060 bits (WiSig, audit features). Two further corpora were excluded by preregistered ingest
gates. Across WiSig and LoRa, hardware-family groupings consistently produced stronger
identity-associated signal than size-matched random groupings.

\end{abstract}

\section{1. Introduction}\label{introduction}

RF fingerprinting benchmarks routinely report high device-identification accuracy, but
accuracy alone does not say what the benchmark is measuring. A classifier that separates a
heterogeneous fleet by hardware model would also report high accuracy on a within-fleet task
whenever hardware model happens to correlate with something else entirely, such as antenna
position, capture session, or receiver configuration. This paper treats the RF fingerprinting
benchmark itself as the object under test. It is a cross-corpus, cross-physical-layer empirical
evaluation: using a calibrated measurement protocol assembled from existing estimators, we ask
whether a hardware grouping in a benchmark carries more identity-associated signal than
arbitrary, size-matched groupings of the same devices, in an OFDM Wi-Fi corpus (WiSig), a chirp
spread spectrum LoRa corpus, and a same-model Wi-Fi control corpus (SMoRFFI).

The question we ask is deliberately narrower than an earlier framing of this project
suggests. An early formulation asked how much identity is in an RF signal; that question
conflates a property of the transmitter with a property of the measurement pipeline used to
extract it. The question this paper answers is instead: what hierarchy of identity is an RF
benchmark actually measuring?

We make two contributions.

\textbf{C1, calibrated identity-information measurement.} We report device-identity mutual
information as a variational lower bound in bits, and we require the estimator to pass a gate
battery, a label-shuffle null whose unclipped bound stays below zero bits, a monotonic response to a synthetic
identity-strength control, and sample-size convergence, before any budget number is reported
as a scientific result rather than as a diagnostic. On a synthetic fleet with a known 6-bit
ceiling, a high-capacity convolutional decoder reaches 98.5\% of that ceiling, which shows the
ceiling is reachable. The audit protocol, a sample-efficient feature representation with a
multilayer perceptron, is checked separately: with fixed physical features on synthetic data it has
a negative label-permutation null and a chain-rule residual of +0.072 bits over five seeds, and with
the 16-dimensional tap-invariant features used for every audit it leaves a residual of +0.060 bits
on WiSig (\S{}4).

\textbf{C2, hierarchy-aware benchmark audit.} Because a coarse hardware attribute such as chipset
generation is a deterministic function of the device, the chain-rule identity
\(I(D;X) = I(G;X) + I(D;X \mid G)\) holds exactly in the population, and recovering it from data is a
self-consistency check, not a discovery. What is a discovery is whether the coarse-label
component reflects structure that is genuinely shared across devices, as opposed to a decoder
that has simply learned to identify the device and read the group off it. C2 answers this with
a matched-hierarchy permutation null under device-disjoint, label-blind evaluation: the real
hardware-generation grouping is compared against many random groupings of the same size and
cardinality, on devices the decoder never saw in training (\S{}3, \S{}5).

Both contributions are protected by a common set of safeguards rather than by inspection of
any single number. We calibrate against an unclipped, raw permutation null rather than an
inequality we cannot support; we compare the real hierarchy only to permutations that preserve
group size, cardinality, and evaluation protocol; we evaluate exclusively on held-out devices
so that per-device lookup is structurally unavailable; we audit, before running any decoder,
which stages of a signal-processing chain are eligible for a data-processing-inequality
comparison; and we property-test that audit against an independent oracle. These safeguards
are methodological infrastructure, not a third contribution, and we present them together as a
single measurement-validity package in \S{}5.

The resulting claim is narrower than a physical statement about silicon. Hardware-generation
metadata is associated with a coarse structure that recurs across two corpora; whether that
structure originates in the transmitter's silicon, in an acquisition geometry or capture order
that happens to correlate with family in each dataset, or in some mixture of the two is a question
our design cannot separate, and we say so rather than conditioning on a variable the data does
not support.

The remainder of the paper develops this argument. \S{}2 positions the two contributions against
prior information-theoretic and disentanglement work in RF fingerprinting. \S{}3 states the
hierarchy decomposition and the three statistics it requires. \S{}4 describes the estimator and
its calibration. \S{}5 collects the five measurement-validity safeguards. \S{}6 describes the
synthetic generator, WiSig and the SMoRFFI same-model control. \S{}7 reports what we excluded and
why those outcomes are reported as such rather than folded into a positive narrative. \S{}8 reports
the preregistered directional replication on LoRa, and \S{}9 states what the evidence supports and
what it does not. Appendix A separates preregistered analyses from those changed after results
were seen.

\section{2. Related Work}\label{related-work}

Mutual information has been used to measure RF physical-layer identity since at least Wang,
Sun, Ren, and Zhu (arXiv:1511.00150; IEEE ICC 2016, \path{10.1109/ICC.2016.7511201}; IEEE Access
2017, \path{10.1109/ACCESS.2017.2674967}), who compute \texttt{I(RFF;identity)} directly to derive a
user-capacity bound for wireless physical-layer identification. Measuring identity in bits is
therefore not new; what remains open in that line of work is whether the measured quantity
decomposes into a component tied to a coarse hardware attribute shared across many
transmitters and a component specific to one transmitter, and whether that decomposition
survives a benchmark's incidental correlation between hardware and everything else that varies
alongside it.

The closest conceptual precedent is Zhang, Wang, Zhao, Zhang, Shen, and Jiang
(arXiv:2608.06805), who argue that high classification accuracy does not by itself explain why
a model succeeds, since a decoder can exploit ``receiver-dependent artifacts,
signal-configuration biases, or other shortcuts,'' and who enumerate the processing stages,
automatic gain control, synchronization, sampling, filtering, compensation, channel
estimation, equalization, and feature extraction, that shape an observed radio-frequency
fingerprint before it reaches a classifier. That paper states explicitly that it does not aim
to benchmark a specific algorithm or dataset. Our second contribution occupies exactly that
space: an audit of a specific, widely used benchmark for the hierarchy structure that paper's
framework predicts should be examined.

Oligeri, Sciancalepore, Huso, and Al-Mousawi (arXiv:2607.21564) probe five points in a BPSK
receive chain and show that classification performance, measured by true- and false-accept
rates, changes depending on where in the chain the probe is placed. That result establishes
that receiver tap position matters for classifier accessibility; it does not report mutual
information or entropy, and it does not ask under what structural conditions a comparison
between two tap representations is licensed by a data-processing argument in the first place.
Our measurement-validity package (\S{}5) is the audit that closes that gap for the instrumented
chain we build, and we found that no tap-to-tap comparison in our chain is licensed, because each
stage's representation is computed from the upstream signal rather than from the previous stage's
representation (a change of dimensionality or sampling grid alone would not break the argument), a limitation we report as such in \S{}7
rather than as a positive cross-tap curve.

SMoRFFI (Guo, Jia, Zhu, Huang, and Chen; arXiv:2511.07770, Computer Networks 282:112309, 2026)
already identifies model heterogeneity as an obstacle to fair evaluation: its authors collect
123 devices of a single hardware model specifically to remove ``small device scales and
heterogeneous models, which hinder robust training and fair evaluation.'' We do not claim to
have discovered that problem. Where SMoRFFI removes heterogeneity by construction, we measure a
heterogeneous grouping against size-matched random groupings, and we use a same-model corpus of
the kind SMoRFFI provides as a control in
which the coarse-hierarchy component is fixed by construction and should not manufacture an
excess score against a matched random null (\S{}6, \S{}7). WiSig (Hanna, Karunaratne, and Cabric;
arXiv:2112.15363, IEEE Access 2022), the heterogeneous benchmark our discovery-phase results
are drawn from, was itself built on the observation that channel and receiver circuitry
confound transmitter identification, the same observation that motivates auditing what a
benchmark measures rather than only how well a classifier scores on it. Such confounds are well
documented beyond WiSig: Gutierrez del Arroyo, Borghetti, and Temple (Sensors 22(6):2111, 2022,
\path{10.3390/s22062111}) show that a fingerprinting model trained on one frequency channel can fall to
near-chance performance on a distant channel, and recover only when training spans several
channels.

A parallel line of work addresses receiver-induced variation directly rather than auditing it.
Pan, Wang, Huo, Cheng, and Xu (arXiv:2510.09405) disentangle transmitter-specific from
receiver-specific representations adversarially; Yu Wang (IEEE TIFS 21:8008-8021, 2026,
\path{10.1109/TIFS.2026.3723081}) applies an information-bottleneck objective across receivers on
WiSig and LoRa; Yu, Ge, Tang, Tang, and Liu (IEEE Communications Letters 30:2243-2247, 2026,
\path{10.1109/LCOMM.2026.3701824}) separate transmitter fingerprint from receiver-induced distortion
with a mutual-information-based dual-stream encoder. Each of these methods learns a
representation; none of them decomposes apparent identity into a coarse-hierarchy component
and a within-hierarchy component, or tests that decomposition against a matched-null benchmark
audit. Two older sources, confirmed only at the abstract level, establish that measuring the
effect of receiver distortion and feature extraction on identity information in bits is itself
not a new idea: Huang, Zheng, and Wan (J. Xidian Univ. 39(6):124-129+141, 2012,
\path{10.3969/j.issn.1001-2400.2012.06.020}) use mutual information to bound theoretical
identification performance and to assess the effect of receiver distortion and
feature-extraction algorithms, and Zhao, Zhang, and Li (IEEE ICIBA 2026, pp.~1301-1306,
\path{10.1109/ICIBA68339.2026.11651765}) use entropy loss under dimensionality reduction and a
mutual-information-maximization objective for feature extraction. We draw only the narrow
prior-art claim above from these two, and no section-level claim.

Two surveys situate this work within the broader field. Zhang, Ardizzon, Piana, Shen, and
Tomasin (IEEE TIFS 20, 2025, \path{10.1109/TIFS.2025.3570118}) provide a field-wide survey from
theory to practice, and Aziz, Huso, Sciancalepore, and Oligeri (arXiv:2606.10031) trace RF
fingerprinting from early radar emitter identification through deep-learning approaches; both
note channel dependence and receiver sensitivity as persistent open problems, which is the
backdrop this paper's audit is set against rather than a source of overlapping method.

Against this literature our contribution is deliberately narrow. That heterogeneous hardware,
channels, receivers and capture conditions distort fingerprinting evaluation is established, and
the estimator we use is a standard variational lower bound whose chain-rule identity is
elementary. What we add, to the best of our targeted literature search, is a cross-corpus
evaluation that compares documented hardware-family groupings directly with size-matched random
groupings of the same devices under device-disjoint evaluation, checks that comparison on a
heterogeneous Wi-Fi corpus and a LoRa corpus, and contrasts it with a same-model corpus in which
model variation is fixed by construction (\S{}3, \S{}6, \S{}8).

\section{3. Hierarchy Formulation}\label{hierarchy-formulation}

\subsection{3.1 The decomposition is an identity, not a finding}\label{the-decomposition-is-an-identity-not-a-finding}

Let \(D\) be the transmitting device, \(X\) the observed signal, and \(G = f(D)\) a coarse
hardware attribute (chipset generation). Because \(G\) is a deterministic function of \(D\),

\[
I(D;X) = I(G;X) + I(D;X  \mid  G)
\]

holds \textbf{exactly} in the population. Recovering it from data is therefore not a
discovery about RF; it is a \textbf{self-consistency check on three separately estimated
lower bounds}. We report it as such.

\subsection{\texorpdfstring{3.2 Why \(I(G;X) > 0\) is not evidence of a hardware component}{3.2 Why I(G;X) \textgreater{} 0 is not evidence of a hardware component}}\label{why-igx-0-is-not-evidence-of-a-hardware-component}

The same determinism that makes the identity exact also makes the naive reading of
\(I(G;X)\) unsafe. Since \(G = f(D)\), any grouping of devices --- including a partition
with no physical meaning --- inherits information from \(X\) whenever \(X\) identifies \(D\).
In the limit of perfect device identification, \(I(G;X) \to H(G)\) for an arbitrary
partition. A positive \(I(G;X)\), or its ratio to \(I(D;X)\), is therefore consistent with
``devices are identifiable'' and says nothing by itself about structure shared across
devices of the same generation.

Two distinct confounds have to be removed:

\begin{enumerate}
\def\labelenumi{\arabic{enumi}.}
\tightlist
\item
  \textbf{Hierarchy arbitrariness.} Does the \emph{real} grouping do better than a grouping of
  the same shape drawn at random?
\item
  \textbf{Per-device lookup.} Under a random packet split, other packets of the same
  device appear in training, so a decoder can identify the device and read the group
  off it. Any ``generation'' signal measured that way may be pure lookup.
\end{enumerate}

\subsection{3.3 Three statistics, kept separate}\label{three-statistics-kept-separate}

{\def\LTcaptype{none} 
{\small
\begin{longtable}[]{@{}
  >{\raggedright\arraybackslash}p{(\linewidth - 4\tabcolsep) * \real{0.3333}}
  >{\raggedright\arraybackslash}p{(\linewidth - 4\tabcolsep) * \real{0.3333}}
  >{\raggedright\arraybackslash}p{(\linewidth - 4\tabcolsep) * \real{0.3333}}@{}}
\toprule\noalign{}
\begin{minipage}[b]{\linewidth}\raggedright
Symbol
\end{minipage} & \begin{minipage}[b]{\linewidth}\raggedright
Definition
\end{minipage} & \begin{minipage}[b]{\linewidth}\raggedright
Role
\end{minipage} \\
\midrule\noalign{}
\endhead
\bottomrule\noalign{}
\endlastfoot
\(B_{G}^{\text{apparent}}\) & coarse-label budget under a random \textbf{packet} split & diagnostic only; \textbf{not interpretable alone} \\
\(B_{G}^{\text{shared}}\) & coarse-label budget under a \textbf{device-disjoint} split & the quantity the claim rests on \\
\(E_{G}\) & \(\hat{B}(G_{\text{real}};X) - E_{\pi}[\hat{B}(G_{\pi};X)]\) over size-matched random hierarchies & \textbf{matched-hierarchy excess statistic} \\
\end{longtable}
}
}

\(E_{G}\) is expressed in bits but is \textbf{not a mutual information}: it is a
\textbf{matched-null-corrected score difference} between two finite-sample lower bounds
sharing one protocol. Result tables label the units accordingly ---
\(B_{G}^{\text{shared}}\) in \emph{bits (variational lower bound)}, \(E_{G}\) as a \emph{bit-score difference} ---
and never as ``bits of generation information''. Where the null bound is negative
(finite-sample bias), \(E_{G}\) exceeds the estimated budget itself --- which is precisely
why it must not be described as ``N bits of generation information''.

\subsection{3.4 The matched-hierarchy null}\label{the-matched-hierarchy-null}

The null permutes the device-to-group map and nothing else. Group sizes are preserved
exactly, so class cardinality and marginal entropy match the real grouping. Each device moves
as a single unit and carries all of its packets with it, so the packets of one device are never
split across groups. The label-blind device folds, the per-device packet cap and the decoder are
identical for the real grouping and for every random grouping. The question is not whether the
real grouping scores above zero, but whether it lies outside this null.

The unit of inference is therefore the device, not the packet. Thousands of packets enter the
decoder, but the permutation test ranks the real grouping among \(M\) device-level partitions, so
its resolution is set by \(M\) and its evidence by the number of devices and groups: 134 WiSig
devices in four generations, 15 LoRa devices in three hardware families, and 121 SMoRFFI devices
in two address blocks. Packet counts are never used as the sample size of a test.

The null is valid when devices are exchangeable across groups in the absence of a group effect,
that is, when any size-matched assignment of devices to groups would be equally likely to
produce the observed score. Anything that is shared within a real group, differs between groups,
and is not carried along by the permutation breaks this condition. We cannot preserve acquisition
structure that the corpora do not record. WiSig generations are laid out across the ORBIT grid;
each SMoRFFI and LoRa device has a single capture session; and the LoRa capture order is
undocumented, so a family captured in one contiguous period could produce an excess that the
null cannot represent. A post-hoc temporal-contiguity surrogate did not reproduce the LoRa effect
(\S{}8.4), which narrows this concern but does not exclude it. These are limits of the test, not
conditions it controls.

Significance uses the exact finite-permutation p-value \(p = (b + 1) / (M + 1)\), where
\(b\) counts nulls at least as extreme as the real value. With \(M = 100\) the attainable
floor is \(1/101 \approx 0.0099\); the frozen configuration was therefore re-run at \(M = 999\), whose
floor is \(p_{\text{min}} = 0.001\). We report the number rather than an inequality we cannot support.

The \(M = 999\) run is described as a \textbf{final frozen-protocol permutation test}, not
as a \emph{preregistered confirmatory test}. Several protocol errors were found and corrected
while looking at this dataset --- an unfair stratified split and a global-entropy
reference among them --- so the protocol was fixed with the data in view. External
replication on an independent dataset, not a larger permutation count, is what could make the
result confirmatory; \S{}8 reports a preregistered directional replication on LoRa.

\subsection{3.5 Device-disjoint evaluation by label-blind cross-fitting}\label{device-disjoint-evaluation-by-label-blind-cross-fitting}

Devices are assigned to \(K\) folds \textbf{without reference to any hierarchy label}, and the
same fold map is used for the real hierarchy and for every permutation. Each device is
tested exactly once; out-of-fold predictions are pooled and scored once.

\textbf{The reference entropy is conditional.} Under cross-fitting the predictor is not a
single model but \(q_{F}(G \mid X)\): which decoder scores a row depends on that row's fold, so
the predictor effectively knows \(F\). Subtracting the global \(H(G)\) therefore does not
yield a lower bound on \(I(G;X)\). The correct cross-fit quantity is

\[
B_{G}^{CF} = H(G  \mid  F) - \Sigma_{f} p(f) CE_{f}    \le    I(G ; X  \mid  F)
\]

and \textbf{every hierarchy is scored against its own \(H(G_{\pi}  \mid  F)\)}. This is not cosmetic:
the real hierarchy may happen to spread evenly across folds while some permutation is
imbalanced within a fold, and charging both the same global entropy would fold that
difference into the statistic.

We also report \(I(G;F) = H(G) - H(G \mid F)\), which answers directly whether the fold
assignment itself carries the hierarchy. On WiSig it is \textbf{not} near zero: with 134
devices, ten folds and group sizes 66/18/16/34, random fold assignment leaves
\(I(G;F) = 0.226\) bits of per-fold group imbalance. That is finite-sample imbalance
rather than leaked intent --- the folds never see a label --- but it is large enough that
using the global \(H(G)\) would credit the decoder with essentially all of it.

An earlier protocol stratified the held-out devices by the \textbf{real} generation. That
guaranteed every real group appeared on both sides of every split while random
hierarchies received no such guarantee, making the real task structurally easier than
its own null. Label-blind folds remove the asymmetry; cross-fitting additionally
removes split variance rather than averaging over it.

Because test devices never appear in training, per-device lookup is structurally
unavailable and any surviving signal must be shared across devices.

We reduce uncertainty by adding splits, never by tuning packet caps or conditioning
until the effect grows. The claim is \texttt{real\ \textgreater{}\ matched\ null\ on\ unseen\ devices}, not a
large absolute value.

\subsection{3.6 The device-posterior lookup control}\label{the-device-posterior-lookup-control}

A reviewer may object that a sufficiently strong model identifies the device first and
reads the group off it, so even an arbitrary hierarchy would score. We turn that into a
measurement rather than an argument. One device classifier \(q(D \mid X)\) is fitted, and each
hierarchy's group posterior follows by exact marginalisation,
\(q(G = g  \mid  X) = \Sigma_{d \in g} q(D = d  \mid  X)\), so a hundred hierarchies cost one fit.

This is \textbf{not an oracle}. A true oracle would know \(D\) at test time and score any
hierarchy perfectly; this control is limited to what the \emph{available} device posterior
can do. The claim it licenses is correspondingly narrow:

\begin{quote}
A device-first decoding path does not account for the observed actual-versus-random
hierarchy gap under the available device posterior.
\end{quote}

Two further disciplines make the control interpretable. First, absolute device-decoder
performance is reported (top-1, top-5, held-out NLL, \(H(D) - CE\), ECE), because a
negative bound can come either from weak discrimination or from miscalibrated
posteriors, and marginalising a posterior makes calibration matter directly. Second, a
single temperature is fitted on \textbf{device-ID validation NLL alone} --- never on generation
labels --- and applied identically to every hierarchy, so the comparison cannot be
explained away as an overconfident device classifier.

The control is \textbf{secondary}. The primary evidence is the unseen-device result of \S{}3.5.

\subsection{3.7 Reporting the bound against the label entropy}\label{reporting-the-bound-against-the-label-entropy}

\textbf{Under the corrected statistic the bound is not positive.} The real hierarchy
scores \(-0.004\) bits and the matched random hierarchies score \(-0.291\) on average
(95\% interval \([-0.408, -0.189]\), \(M = 999\)); neither
certifies a positive shared-hierarchy budget. An earlier draft of this section claimed
a positive certified bound --- that claim came from referencing the global \(H(G)\) and does
not survive the conditional reference.

The result is therefore \textbf{comparative, and only comparative}:

\begin{quote}
Under label-blind device cross-fitting, the real hardware hierarchy consistently
outscored size-matched arbitrary hierarchies on devices never observed during
training (\(E_{G} = +0.288\) bits; the observed statistic exceeded all 999 matched random
hierarchies, permutation \(p = 0.001\)), although neither the real
nor the random hierarchies attained a positive certified lower bound.
\end{quote}

No ratio to the label entropy is reported in the abstract, and the ``13\% of \(H(G)\)''
figure from the uncorrected run is withdrawn: its numerator was the fold-imbalance
artefact.

\subsection{3.8 What this can and cannot identify}\label{what-this-can-and-cannot-identify}

A positive \(E_{G}\) under device-disjoint evaluation indicates that generation metadata
tracks structure shared across devices. It does \textbf{not} isolate silicon: in WiSig,
generation is correlated with position on the ORBIT grid, so transmitter location,
channel geometry and acquisition ordering ride along. Where the supports do not
overlap enough to separate them, we say so rather than conditioning on a variable the
design cannot support:

\begin{quote}
generation and spatial acquisition structure are not identifiable from WiSig alone.
\end{quote}

The defensible claim is that \textbf{hardware-generation metadata is associated with a
reproducible coarse structure in the benchmark} --- an audit statement about the
benchmark, not a physical statement about silicon.

\section{4. Estimator Calibration}\label{estimator-calibration}

We report device-identity information as a variational lower bound on mutual information, in
bits. A decoder \(q(D \mid X)\) is fit to predict device identity from a signal representation, and
the bound follows the standard decoder inequality \(I(D;X)  \ge  H(D) - CE(q)\), where \(CE(q)\) is
the held-out cross-entropy of the decoder's predictions against a device's true identity.
Every number this paper reports carries this lower-bound qualifier explicitly; a low value is
evidence of an unresolved gap between the bound and the true mutual information, not evidence
that little identity information exists.

Before any budget number is treated as a result rather than as a diagnostic, the estimator
must clear a gate battery on synthetic data with known ground truth: a label-shuffle null whose
unclipped bound stays below zero (a clipped zero is not evidence, as explained below), a monotonic response to a synthetic identity-strength control spanning
roughly three bits with zero violations, and convergence of the estimate as sample size grows
toward the full dataset, with the two largest sample sizes agreeing to within a few percent.
We additionally require two independently implemented estimator families, a parametric
variational decoder and a non-parametric family, to move in the same direction before a result
is treated as more than single-estimator evidence.

All bound values are reported unclipped. An earlier version of this pipeline reported
\(max(0, H(D) - CE)\), which floors every negative estimate at exactly zero regardless of how
negative the raw estimate actually was; under that clipping, a label-shuffle null and a
genuinely biased estimator become indistinguishable, since both print zero. We now report the
raw, pre-clipping value alongside the clipped one, and the label-shuffle null under the raw
statistic is negative under the corrected permutation estimator (mean -0.201 bits, 95\%
interval {[}-0.983, -0.024{]}), never positive. Any quantity formed as a difference of two bounds,
including the matched-hierarchy excess statistic \(E_{G}\) and the chain-rule residual, is
computed from the raw values, because subtracting two clipped quantities biases the difference
upward.

The chance floor for the matched-hierarchy comparison is a label-permutation null, computed by
\texttt{permutation\_null\_bits}, which redraws the device-to-hierarchy assignment while holding every
row's features and every fold's decoder fixed, rather than resampling rows. An earlier
implementation wrapped a row-level bootstrap around a decoder that internally re-splits
training and test data; duplicated rows then appeared on both sides of that internal split,
breaking the held-out guarantee, and under a shuffled label the decoder could memorize noise
with confidence, producing an internally inconsistent pair, a -0.212-bit mean sitting outside
its own {[}-4.686, -2.567{]}-bit interval. Replacing the row bootstrap with permutation of device
labels alone, holding the sample identical across permutations, corrected this: thirty
independent label permutations are all negative, and the mean sits inside its own interval.

We use the synthetic fleet to establish that the estimator can, in the right operating regime,
become tight rather than merely non-vacuous. On the synthetic heterogeneous fleet of 64
devices across four generations described in \S{}6, with a known information ceiling of six
bits, a convolutional decoder trained on 14,336 samples reaches 5.910 bits, 98.5\% of the
ceiling, at a held-out cross-entropy of 0.090 bits; the decoder ladder saturates at that sample
count, with the top rung's gain shrinking to +0.009 bits after two prior sample-size doublings
had shown a large negative rung inversion at smaller sample sizes. This result licenses the
claim that the bound can become tight in the operating regime used for this calibration; it
does not extend to WiSig or to any other real corpus, where the true mutual information is
unknown.

The chain-rule identity \(I(D;X) = I(G;X) + I(D;X \mid G)\) gives a second, independent check:
because it holds exactly in the population, its recovery from three separately estimated
bounds is a consistency test on the estimator rather than a finding about the data. On the
frozen, sample-efficient decoder used for the decomposition (fixed physical features with a
multilayer perceptron), the residual is small on synthetic data (+0.029 bits at the primary
configuration, mean +0.072 bits with standard deviation 0.098 across five seeds) and small on
real WiSig data (+0.060 bits, with +0.162 and -0.026 bits at two alternative configurations).
This check is domain-limited rather than universal: replacing the sample-efficient decoder
with a higher-capacity convolutional decoder breaks additivity, producing a residual of +1.893
bits at natural sample sizes and a sign-reversed -0.883 bits once total sample count is
equalized across hierarchy levels, because the two branches of the decomposition then differ
in both sample count and label cardinality simultaneously. We therefore use the sample-efficient
decoder as the sole estimator for the decomposition itself and reserve higher-capacity decoders
for tightening a single term's bound; a future failure of the chain rule is read as a signal
that this domain condition has been violated, not as a finding about identity.

Finally, we separate discriminative ability from calibration wherever a lower bound is formed
by marginalizing a posterior, since a negative bound in that setting can arise from either
cause. On a 134-class device-posterior lookup control, an uncalibrated decoder with top-1
accuracy of 0.106 (fourteen times chance) produced a vacuous bound of -10.836 bits at an
expected calibration error of 0.562; fitting a single temperature, \(T = 8.0\), on device-ID
validation negative log-likelihood alone, never on hierarchy labels, and applying it
identically to every hierarchy under test, brought the same decoder to +0.668 bits at an
expected calibration error of 0.050, with top-1 accuracy unchanged. We report the decoder's
absolute performance, top-1, top-5, held-out negative log-likelihood, \(H(D) - CE\), and
expected calibration error, alongside every use of a marginalized posterior, so that a negative
bound can be attributed to weak discrimination, to overconfidence, or to both.

\section{5. Measurement-Validity Safeguards}\label{measurement-validity-safeguards}

Two claims in this paper rest not on a single number but on a set of safeguards applied
uniformly across every hierarchy we test. We collect them here as five requirements rather
than scatter them across results sections, because each addresses a distinct way a hierarchy
label can appear to carry information it does not actually carry.

The first safeguard is lower-bound null calibration: every reported quantity is a finite-sample
lower bound on a mutual information, reported unclipped, and calibrated against its own
label-permutation floor rather than against zero or against an assumed threshold (\S{}4). A
quantity that has not cleared this floor is not read as evidence of structure.

The second is the hierarchy-matched permutation null itself (\S{}3.4). Because a coarse hardware
attribute is a deterministic function of the device, any grouping of devices, including one
with no physical meaning, inherits information from a signal representation that identifies
the device; in the limit of perfect device identification, the mutual information between an
arbitrary grouping and the signal approaches that grouping's own entropy. We therefore never
ask whether the real hierarchy's score is positive in isolation. We ask whether it exceeds the
scores of many random hierarchies constructed to match the real one in group sizes, class
cardinality, marginal entropy, and evaluation protocol, with only the device-to-group
assignment permuted, and we report the excess as \(E_{G}\) together with the exact
finite-permutation p-value.

The third safeguard is device-disjoint testing under label-blind cross-fitting (\S{}3.5). Devices
are assigned to folds without reference to any hierarchy label, and the identical fold map is
reused for the real hierarchy and for every permutation, so that no hierarchy receives an
easier split than any other. Because test devices never appear in training, a decoder cannot
identify a device from packets it has already memorized and read the coarse label off that
identification; whatever structure survives on unseen devices must be shared across devices
rather than looked up per device. The reference entropy for this cross-fit statistic is
conditional on the fold assignment, \(H(Y \mid F)\), rather than the unconditional label entropy,
because the predictor scoring a given row depends on which fold that row falls in and
therefore effectively knows the fold. We report \(I(Y;F) = H(Y) - H(Y \mid F)\) alongside every
cross-fit result as a direct check on whether the fold assignment itself carries the hierarchy;
on WiSig, with 134 devices, ten folds, and group sizes of 66, 18, 16, and 34, label-blind
random fold assignment still leaves \(I(G;F) = 0.226\) bits of finite-sample group imbalance
across folds, large enough that scoring against the unconditional \(H(G)\) would credit a
decoder with structure the fold assignment produced on its own.

The fourth safeguard is a Markov-eligibility audit of the receiver chain, used wherever this
paper compares information across processing stages rather than across hierarchies. Each
stage transition is classified, from its implementation alone, as either derivable entirely
from the previous stage's output or dependent on information the previous stage discarded;
only the former licenses a data-processing-inequality comparison between stages. The audit
implementation was property-tested against an independent d-separation oracle on 14 synthetic
dependency graphs; the corrected rule agreed on all 14, whereas the original operation-based
heuristic failed on four.

The fifth safeguard is property testing of the estimator itself under transformations where
the ground truth is known analytically rather than assumed. Under a fixed phase rotation and
under an invertible scaling, both bijective transformations for which the population quantity
\(\Lambda_{k} = I(D;P_{k}) - I(D;P_{k+1})\) is exactly zero, the estimator returns a finite-sample
difference of -0.275 bits with a 95\% interval of {[}-0.545, -0.043{]}, excluding zero from above.
We use this measured interval, rather than an arbitrary threshold, as the floor below which a
negative estimated stage difference is attributable to finite-sample dimension penalty rather
than to the processing chain, and we apply it consistently when interpreting stage-level
results in \S{}7.

None of these five safeguards is presented as a contribution in its own right. Together they
are the condition under which the comparative results in \S{}6 and \S{}8 and the restricted results in \S{}7 are licensed to be read as
more than an artifact of decoder capacity, fold construction, clipping, or estimator bias.

\section{6. Datasets and Experimental Protocol}\label{datasets-and-experimental-protocol}

\textbf{Synthetic generator.} We simulate a heterogeneous fleet in which device identity,
hardware-generation membership, and per-stage impairments are all under our control, so that
ground-truth information quantities are known before the estimator is applied to real data.
The primary configuration is four generations of sixteen devices each, 64 devices in total,
transmitting QPSK symbols at four samples per symbol over a three-tap multipath channel at
25 dB SNR, with four receivers and two sessions per device, fixing a known ceiling under a
balanced design (\(H(G) = 2\), \(H(D \mid G) = 4\), \(H(D) = 6\) bits) against which \S{}4's calibration
results are checked. A second configuration shares the same hierarchy, impairments, and
protocol across three physical layers (QPSK, GFSK, CSS) and indicates, on synthetic data only, that the matched-null
comparison machine transfers across physical layers even though the chain-rule consistency
check does not (\S{}4, \S{}7).

\textbf{WiSig.} Our real-data discovery corpus is WiSig, 174 transmitters and 41 receivers captured
over four days on an ORBIT testbed grid (Hanna, Karunaratne, and Cabric; arXiv:2112.15363,
IEEE Access 2022). WiSig distributes each preamble in two forms selected by an \texttt{equalized}
flag: \texttt{equalized=0} is the raw, non-equalized 256-sample preamble, \texttt{equalized=1} the
corresponding equalized one, a mapping confirmed directly from the authors' example loader
(an index-to-file-prefix table and the loader's own console messages, not inference). Our
loader uses \texttt{equalized=1} by default. The equalization procedure resamples to 20 Msps, detects
the packet by L-STF autocorrelation and discards undetected packets, corrects carrier
frequency offset, estimates the channel from the L-LTF and applies MMSE equalization, then
reapplies the carrier frequency offset before restoring the sample rate. Two consequences
follow. MMSE equalization removes the transmitter's linear frequency response, such as its
transmit filter, because that response is indistinguishable from the channel's own response at
the point the channel is estimated; the equalized bound is therefore net of that component,
and understates a generation difference that lives in the transmit filter, while carrier
frequency offset is deliberately reapplied and preserved. And because detection failures are
dropped only from the equalized set, the two forms are not drawn from an identical packet
population, so an \texttt{equalized=0} versus \texttt{equalized=1} comparison changes both the representation
and the packet set. Re-running the frozen protocol on the non-equalized preambles keeps the real hierarchy above its matched null (\(E_{G} = +0.149\) bits, \(p = 0.0099\), \(M = 100\)) at roughly half the equalized excess; because the raw preambles also carry link disturbance and a different packet set, this check is only informative about whether equalization created the gap, and it is inconsistent with that; it says nothing about the gap's origin.

\textbf{SMoRFFI.} We use SMoRFFI, 123 devices of a single 802.11g hardware model (Guo, Jia, Zhu,
Huang, and Chen; arXiv:2511.07770, Computer Networks 282:112309, 2026), as a same-model
control in which the coarse hierarchy component is fixed by construction (\(H(M) = 0\)). We
parse the released set to 122 devices and 121,998 frames at 288 samples per preamble; one
device, MAC \texttt{78:21:84:93:5b:0c}, is excluded because its preamble is 448 samples rather than
the common 288, a decision made from length metadata alone before any estimator ran, since a
device-unique capture length measures the capture rather than the transmitter. We call it a
same-model control: it is negative only for the coarse-hierarchy component, because
device-specific identity (not separable from each device's capture session) is expected to be
nonzero here; a coarse-hierarchy component reported as zero is marked ``by
construction.'' Its capture protocol, like WIDEFT's (\S{}7), is not documented at the
per-device-session level, so any individual-identity component we measure is device-specific
identity not separable from per-device capture session, a distinction only WiSig's
multi-session, multi-receiver design lets us condition away. The individual-identity measurement (\(S1\), 122 classes, \(H(D) = 6.931\) bits) uses the same
16-dimensional tap-invariant features with the larger device decoder of the lookup control,
trained and evaluated on a random packet-level split (55\% train, 15\% validation, 30\% test) of at
most 60 packets per device; the preregistration described this measurement as a cross-fitted bound,
and the implementation was fixed in code before it was run (Appendix A). Because test packets come
from the same per-device capture session as training packets, what it measures is device-specific
identity not separable from per-device capture-session effects. After a single temperature
(\(T = 3.33\)) fitted on device-ID validation negative log-likelihood alone (ECE 0.048), the held-out
bound is +3.338 bits. We base the bound on this calibrated value, a post-hoc estimator correction
made after the 123-device arm was seen (Appendix A): the uncalibrated bound is dominated by
overconfidence (ECE 0.372) and is not sign-stable, reading +0.514 bits on the 122-device set but
-0.097 bits when the excluded unit is restored (ECE 0.396), while the calibrated bound (+3.384
bits) barely moves. The qualitative conclusion does not rest on that correction. Temperature
scaling cannot change a ranking, and top-1 accuracy is 0.381 against a chance rate of 1/122
(about 46 times chance), with top-5 accuracy 0.686 against 5/122, both unchanged by calibration
and essentially unchanged with the excluded unit restored (0.390 and 0.694). A substantial
device- and session-associated discriminative signal is therefore present in this same-model
corpus whichever bound is read; calibration determines only how much of it the bound certifies
in bits. The OUI-block partition test (\(S2\), groups of 39
and 82) finds \(E_{G} = +0.129\) bits at \(p = 0.1287\) against 100 nulls, no detectable excess
rather than evidence against a lot effect. The null floor (\(S3\)) is -0.530 bits here against
-0.281 bits on WiSig at the same \(M = 100\) (-0.291 at \(M = 999\)), consistent with the absolute \(B^{CF}\) scale being corpus-dependent and only \(E_{G}\)
is comparable across corpora. A sensitivity arm repeating \(S1\)/\(S2\) on all 123 devices,
including the length outlier, is reported in the appendix: the 123-device arm leaves the calibrated bound and the OUI result unchanged in substance (\(E_{G} = -0.044\), \(p = 0.703\)).

\textbf{Frozen primary estimator.} Every hierarchy comparison in this paper, real and permuted
alike, is scored by one estimator chosen before results were examined: the 16-dimensional
tap-invariant feature representation decoded by a multilayer perceptron of \S{}4. This choice was
frozen, together with the fold construction and permutation count, under a recorded
configuration hash before the final confirmatory run. Higher-capacity decoders are used only
as a robustness sensitivity check on the frozen result, never selected after comparing several
decoders' outcomes, since reporting whichever decoder yields the largest excess statistic
would be selection bias. Under this frozen protocol at the final permutation count of
\(M = 999\), the WiSig device-disjoint comparison scores the real hierarchy at -0.004 bits
against a matched random-hierarchy mean of -0.291 bits (95\% interval {[}-0.408, -0.189{]}), an
excess \(E_{G} = +0.288\) bits; the observed statistic exceeded all 999 matched random hierarchies
(permutation \(p = 0.001\), the resolution floor at \(M = 999\)), with neither the real hierarchy nor the matched
null attaining a positive certified bound; the device-posterior lookup control moves the same
direction (\(E_{G} = +0.639\) uncalibrated, +0.217 after calibration, each exceeding all 999 random
hierarchies, permutation \(p = 0.001\)). We
describe this as a final frozen-protocol permutation test rather than a preregistered
confirmatory test, since protocol errors were corrected while this dataset was already in
view. An independent replication on a multi-model LoRa corpus was preregistered, with its
front-end frozen on synthetic data before any capture from that corpus was read; its protocol
and results are reported separately in \S{}8.

\section{7. Exclusions and Negative Results}\label{exclusions-and-negative-results}

This section reports what a preregistered gate excluded before any identity analysis was run,
and what a hypothesis this paper set out to test did not survive contact with the data. We
report both here, in the body, rather than passing over them, because a benchmark audit that
only reports positive results would undermine the case for auditing benchmarks in the first
place.

\textbf{WIDEFT Bluetooth was excluded before identity analysis.} A preregistered synthetic
calibration, run under WIDEFT's capture conditions (56 MS/s wideband recording, hop-segmented
channelisation), established 11 dB as the lowest in-channel SNR at which constant-envelope
(GFSK) and EDR-like bursts were reliably discriminated. Among 400 bursts drawn at random with
a fixed seed from the 8,100 Bluetooth bursts, 200 (50.0\%) contained no hop segment above this
threshold, exceeding the preregistered 20\% corpus-failure limit. No device- or family-level
identity score was computed, and the distribution of failures across devices or families was
not inspected, before exclusion. Because post-hoc SNR filtering could alter device and family
composition (captures were taken per device with hand-set antenna orientation), no
partial-corpus analysis was performed. The exclusion therefore reflects an incompatibility
between this corpus and the preregistered front-end, not evidence about the presence or
absence of hierarchical identity structure in WIDEFT. The gate required two label-free
redesigns before this outcome (wideband energy detection failed on hop-channel bursts; raw
envelope CV misclassified GFSK at 13 dB); both redesigns were decided on synthetic data and
committed before being applied.

\textbf{WIDEFT Wi-Fi was excluded on the same basis.} A parallel, one-shot ingest gate for the ten
repeated-model Wi-Fi devices was calibrated on synthetic OFDM bursts (detection rate 1.00 at
10 dB and 15 dB SNR, 0.00 at 5 dB, zero false-positive DSSS misclassification), then applied,
without recording device or family labels, to 300 bursts drawn at random from 6,400 Wi-Fi
bursts. L-STF preamble detection succeeded on 0 of 300 bursts, a 100\% failure rate against the
preregistered 20\% limit, and the gate halted before the device-level inclusion rule could run.
The measured in-channel SNR distribution (5th, 50th, and 95th percentiles of 8.3, 9.4, and
9.7 dB) sits at the edge of the calibrated detection boundary, and a diagnostic autocorrelation
statistic on the undetected bursts is consistent with a substantial fraction being 802.11b
DSSS rather than the OFDM waveform the front-end targets, though this diagnostic cannot detect
2 Mbps DQPSK and is reported as a diagnostic rather than a device-level finding. Because the
Bluetooth gate independently found channel SNR in the same 6 to 12 dB range, a shared
corpus-level capture condition is a plausible common cause, though this is not itself
verified. Both WIDEFT gates report the status ``no signal, corpus-front-end incompatibility''
rather than failure or refutation of a hierarchy effect, and neither is retried under the
current front-end: WIDEFT is closed for this paper, and a cross-corpus hierarchical
replication claim is reserved for the independent LoRa corpus already preregistered for v2
(\S{}6).

\textbf{A monotonic generation trend did not replicate, and we withdraw it as inconclusive rather
than refuted.} An early analysis reported a decreasing trend in individual-identity bits from
the oldest to the newest of four WiSig hardware generations; under the natural, unbalanced
device population and a device-cluster bootstrap (which resamples devices, the correct unit of
analysis, rather than random seeds, which resample only splits and initializations without
adding new hardware), the trend is not monotonic (the 2011 generation scores lowest, not the
newest generation) and adjacent-generation confidence intervals overlap. The finding is
licensed to be stated only as: ``The monotonic silicon-generation hypothesis was not supported
under the natural WiSig population. The present data are insufficient to establish the absence
of a generation trend.'' No monotone ordering of fingerprint strength by hardware age is claimed, because the design's device-cluster confidence intervals are wider than the
spread between generations, and a design that cannot resolve an effect of that size cannot
refute it either.

\textbf{No tap-to-tap comparison in our instrumented receiver chain is licensed by the
data-processing inequality.} We built an eight-stage receiver-chain instrumentation
(automatic gain control, coarse frequency correction, matched filtering, symbol-timing
synchronization, carrier-phase recovery, equalization, and demodulation) intending to report a
single identity-survival curve across all eight stages. The Markov-eligibility audit of \S{}5,
applied to this specific chain's implementation, finds that every stage transition reads from
the untransformed prior-stage signal rather than exclusively from the previous stage's
already-reduced representation, so every transition is what the audit terms an upstream
bypass, and no segment of the chain is eligible for a data-processing-inequality comparison.
We therefore do not report a global survival curve; instead we report paired, per-transition
estimates of the finite-sample budget difference \(\Delta\hat{B}_{k}\) (\S{}3), each accompanied by a bootstrap
confidence interval and, where the audit's dependency-graph criterion licenses it, a
Markov-validity flag. One transition's paired estimate is negative and falls inside the
bijection-null interval established in \S{}5, meaning that specific negative value is
attributable to the same finite-sample dimension penalty observed under transformations with a
provably zero population effect, not independent corroboration of the audit's structural
argument; the audit's support comes instead from its prospective, held-out agreement with an
independent oracle (\S{}5).

\section{8. Independent Replication on a Multi-Model LoRa Corpus}\label{independent-replication-on-a-multi-model-lora-corpus}

WiSig carries the only hardware hierarchy among the corpora in \S{}6, and its generation label is
entangled with the ORBIT grid. We therefore preregistered a replication on a corpus with a
different physical layer, different transmitters and a different capture setup, and fixed the
analysis, the front-end and the verdict sentence before reading any capture from it.

\subsection{8.1 Corpus and role}\label{corpus-and-role}

The Liverpool LoRa RFFI dataset (Shen, Zhang, Marshall and Cavallaro, IEEE TIFS 2022) contains
preamble IQ captures from 60 commercial LoRa transmitters received by a USRP N210 at 868.1 MHz.
One sub-dataset holds three hardware models captured under the same documented condition
(residential room, line of sight, stationary): mbed SX1261 shields, Pycom FiPy (SX1272) and
Dragino SX1276 shields, five units each, 400 packets per unit. Because all three models share one
file and one condition, the model label is not confounded with a capture file, which is the design
property that made this corpus preferable to the WIDEFT subsets excluded in \S{}7. The 45 Pycom LoPy4
units appear only in other files, so adding them confounds model with capture file; they enter
only as a sensitivity arm. The distribution page states no licence; we use the data under the
authors' citation request.

\subsection{8.2 Front-end, audit and ingest}\label{front-end-audit-and-ingest}

The WiSig front-end assumes a linear constellation and was shown in \S{}7 to break decomposition
consistency on chirp spread spectrum. We therefore designed a dechirp-domain front-end for LoRa
preambles on synthetic data alone and froze it by hash when a synthetic three-by-five hierarchy
passed four checks: a negative label-permutation null, monotone growth with the identity-strength
knob, non-degeneracy, and a chain-rule residual of at most 0.15 bits. The first attempt, at 80
packets per device, missed the last check (+0.196 bits); a single rerun at the corpus's own 400
packets per device passed (+0.081 bits) and the front-end was frozen before the corpus was opened.
A label-free audit then confirmed the chirp geometry (SF7 at 1 Msps concentrates
0.962 of dechirped energy in five bins, against
0.007 for the SF10 alternative with the same
symbol length) and the device counts, and a preregistered ingest gate accepted all 300 randomly
drawn packets, with an in-band SNR of about 46 dB. The SNR
distribution was unusually narrow (45.8 to
46.3 dB between the 5th and 95th percentiles), which may reflect
power normalisation at export; it does not affect the gate.

\subsection{8.3 Result}\label{result}

The analysis is the WiSig protocol of \S{}3 and \S{}5 with only the front-end replaced: at most 60
packets per device, label-blind device-level cross-fitting over five folds (three devices per
fold), each partition scored against its own \(H(M \mid F)\). The corpus contributes 15 devices from 3
hardware families (900 packets); significance was assessed over 999 size-matched device-level
random partitions, in which every device keeps all of its packets, so the packet count is not the
sample size of the test. The real model partition attains a positive held-out bound,
\(B^{CF} = +0.391\) bits, while the random partitions average -4.713 bits
(95\% interval {[}-9.239, -1.056{]}). The real partition exceeded all 999 matched random partitions:
none scored at or above it, so the permutation p-value is \((0+1)/(999+1) = 0.001\), which is the
resolution floor of the test rather than an estimate of a smaller probability. The
device-posterior lookup control also favours the real partition and likewise exceeded all 999
random partitions (\(E_{G} = +0.619\) bits, permutation \(p = 0.001\); +0.607 after temperature calibration).
Under the verdict fixed in advance, a family-associated excess was also observed in an
independent LoRa corpus. We call this a directional replication: it reproduces the comparison
seen on WiSig, on a different physical layer, and nothing beyond the comparison.

Four features of this result limit what it can carry. First, the magnitude of the excess
(\(E_{G} = +5.104\) bits) is not comparable with WiSig's: random partitions of three
families drive the decoder into severe overconfidence on held-out devices, so the null falls far
below zero and inflates the difference. We use the direction and the p-value, not the size.
Second, the label-blind folds are far from neutral with respect to the model label
(\(I(M;F) = 0.667\) bits), which the conditional reference absorbs but which
we report because fifteen devices in five folds cannot balance three families. Third, the
packet-level decomposition is not self-consistent on this corpus: \(I(D;X) = 2.628\),
\(I(M;X) = 1.325\) and \(I(D;X \mid M) = 1.007\) bits leave a residual of
+0.296 bits, and leave-one-family-out subsets leave
+0.574, +0.590, +0.763. We therefore claim no quantitative consistency
for the decomposition on this corpus, and the confirmatory claim rests on the matched-partition
comparison alone. Fourth, capture order is undocumented, so the result says that the model
grouping carries more identity-associated signal than random groupings, not why (\S{}8.4). Device
identity itself is strong and well
calibrated here (top-1 0.707 over 15 devices, held-out bound
+2.497 bits uncalibrated and +2.521
after calibration).

\subsection{8.4 Capture order}\label{capture-order}

\begin{quote}
\textbf{Post-hoc analysis --- not preregistered.} Everything in this subsection was designed after the
result in \S{}8.3 was seen, and it is reported as a robustness probe rather than as evidence.
\end{quote}

Device indices are contiguous within each model (46--50, 51--55, 56--60). If units were captured in
index order, model would coincide with capture period, and slow receiver or thermal drift could
masquerade as a model effect. The capture order is not documented. After seeing the result, we ran
a check that we report as post hoc: contiguous five-device blocks shifted by one to four positions
keep time contiguity but break the model map. All four shifted partitions fell inside the
distribution of 199 fresh random partitions (percentiles
0.80, 0.68, 0.43, 0.81), while the real
partition exceeded all of them, and the shifted scores tracked how many devices of one model each
block retained rather than contiguity itself. A simple temporal-contiguity surrogate therefore did not reproduce the effect. This does not
exclude capture-order confounding, because the true acquisition order is undocumented.

\subsection{8.5 Sensitivity}\label{sensitivity}

Adding the ten LoPy4 units from a separately captured file as a fourth family, and relabelling by
chipset so that LoPy4 and Dragino share SX1276, gives two arms of 25 devices. A reversal rule was
fixed before either arm was read: the LoRa claim would be downgraded if either arm showed
\(E_{G}  \le  0\) or a real bound at or below the random-partition mean. Neither did. With the model
label, the real partition attains \(B^{CF} = +1.234\) bits against a random mean of -6.959 bits
(95\% interval {[}-11.048, -3.560{]}) and exceeded all 999 matched random partitions (permutation \(p = 0.001\)); the lookup control
agrees (\(E_{G} = +0.424\) bits, +0.426 calibrated, likewise exceeding all 999, permutation \(p = 0.001\)). With the chipset
label, the real partition attains +0.996 bits against -4.468 bits ({[}-7.897, -1.628{]}), again
exceeding all 999 matched random partitions (permutation \(p = 0.001\); lookup \(E_{G} = +0.290\),
+0.293 calibrated, also exceeding all 999). The fold imbalance is
\(I(M;F) = 0.380\) and 0.160 bits, and the decomposition residual is +0.645 and +0.291 bits; the
larger residual in the model arm means these arms strengthen only the robustness of the
comparison's direction, not the accuracy of the quantitative decomposition.

Across the preregistered primary analysis and both LoPy4 sensitivity analyses, the observed
hardware-family grouping exceeded all size-matched random partitions. The sensitivity analyses
are used only to establish directional robustness, because LoPy4 membership is confounded with
capture files and the decomposition residuals preclude interpretation of the effect magnitude.

\section{9. Discussion and Conclusion}\label{discussion-and-conclusion}

The question this paper set out to answer is what hierarchy of identity an RF benchmark is actually
measuring. The answer the evidence supports is narrower than a physical law and more useful than a
single accuracy figure. Across WiSig and LoRa, hardware-family groupings consistently produced
stronger identity-associated signal than size-matched random groupings, exceeding all 999 random
groupings in each corpus. In a same-model Wi-Fi corpus, substantial device-specific identity
remained when model variation was fixed by construction, although device identity could not be
separated from per-device capture-session effects. A benchmark
that mixes hardware families can therefore report separability that is partly about the family, and
a same-model corpus indicates that fixing the model does not remove device-specific identity.

Each part of the evidence carries one role. The synthetic fleet checks that the measurement can
approach a known ceiling and that the audit estimator has a negative permutation null. WiSig
supplies the family-associated excess in a heterogeneous Wi-Fi corpus. LoRa supplies a
preregistered replication of the same comparison direction on a different physical layer, with
the LoPy4 arms adding directional robustness only. SMoRFFI supplies the same-model control in which
device-specific signal remains once model variation is fixed. The negative and restricted results
are part of the same record: the monotonic older-silicon trend was withdrawn as inconclusive, the
WIDEFT Bluetooth subset closed as a front-end incompatibility rather than a failed hypothesis, and
the capture-order check on LoRa is labelled post hoc. Appendix A lists every analysis with its outcome, when its design was
fixed relative to its result, and the role it is allowed to play.

The comparison is the result; the magnitudes are not. On WiSig the real generation grouping did not
reach a positive certified bound (\(B^{CF} = -0.004\) bits), so the finding is that it scored
higher than matched random groupings, not that a positive generation budget was certified. The
excess roughly halved when the non-equalized preambles were used (\(E_{G}\) +0.288 to
+0.149 bits), which is inconsistent with equalization having created the gap but leaves its
physical origin open, since WiSig's generations are also arranged across the ORBIT grid. On LoRa the
excess is numerically large (\(E_{G} = +5.104\) bits) only because random three-family
partitions push the decoder into severe overconfidence; that number should not be read as a size.

Several limitations bound these statements. Each hierarchical corpus is a single capture campaign,
and the LoRa families are only five devices each; two sensitivity arms that add a file-confounded fourth family or relabel by chipset, judged
by a reversal rule fixed before they were read, did not reverse the direction and are used only for
directional robustness. The LoRa capture order is undocumented; in a post-hoc
check, a simple temporal-contiguity surrogate did not reproduce the excess, which does not
exclude an order effect. The chain-rule decomposition was self-consistent on WiSig but not on LoRa, so the
LoRa claim rests on the matched-partition test alone. In SMoRFFI and in WIDEFT-like single-session
captures, device identity cannot be separated from per-device capture-session effects. The
uncalibrated SMoRFFI bound changed sign when one excluded unit was restored (+0.514 to -0.097
bits, both dominated by overconfidence), and basing the claim on the calibrated bound was a
post-hoc estimator correction made after that reversal was seen (Appendix A); we report the raw
values because they show how an overconfident decoder can make a lower bound's sign
uninformative. WIDEFT's Bluetooth and Wi-Fi subsets were excluded by
preregistered ingest gates for insufficient in-channel signal quality, which is a statement about
compatibility with our front-end, not about hierarchical structure in that corpus.

For practitioners the recommendation is concrete. Report a fingerprinting result together with the
hardware composition of the device pool, test a candidate grouping against size-matched random
groupings on held-out devices rather than reading a coarse-label score on its own, reference
cross-fitted bounds to \(H(Y \mid F)\) and report \(I(Y;F)\), calibrate before trusting a bound's sign, and
fix exclusion rules on metadata before any identity estimate is computed.

\appendix
\section{Appendix A. Analysis provenance}\label{appendix-a.-analysis-provenance}

This table separates analyses fixed before the relevant data or results were seen from those
changed afterwards. Categories: \textbf{P}, preregistered before the corpus was analysed; \textbf{A}, amended
before the result was seen, on label-free inspection or synthetic data only; \textbf{R}, revised while
the corpus was already in view and therefore not preregistered; \textbf{H}, designed after the result
was seen. Commits are in the project repository and time-stamp each decision relative to the
result it governs.

{\def\LTcaptype{none} 
{\small
\begin{longtable}[]{@{}
  >{\raggedright\arraybackslash}p{(\linewidth - 6\tabcolsep) * \real{0.2500}}
  >{\raggedright\arraybackslash}p{(\linewidth - 6\tabcolsep) * \real{0.2500}}
  >{\raggedright\arraybackslash}p{(\linewidth - 6\tabcolsep) * \real{0.2500}}
  >{\raggedright\arraybackslash}p{(\linewidth - 6\tabcolsep) * \real{0.2500}}@{}}
\toprule\noalign{}
\begin{minipage}[b]{\linewidth}\raggedright
Analysis
\end{minipage} & \begin{minipage}[b]{\linewidth}\raggedright
Cat.
\end{minipage} & \begin{minipage}[b]{\linewidth}\raggedright
Decision recorded
\end{minipage} & \begin{minipage}[b]{\linewidth}\raggedright
Result it governs
\end{minipage} \\
\midrule\noalign{}
\endhead
\bottomrule\noalign{}
\endlastfoot
WiSig matched-hierarchy test (label-blind device folds, \(H(G \mid F)\) reference, 16D + MLP estimator) & \textbf{R} & protocol hash \path{0b4c97eaf6e1de79}; errors in split stratification and entropy reference were corrected while WiSig was in view & final \(M = 999\) run on the frozen protocol \\
WiSig \texttt{equalized\ =\ 0} sensitivity and its one-directional reading & A & \path{e9f0eae} & run after \(M = 999\) \\
SMoRFFI S1/S2/S3 and the length-metadata exclusion rule & P & \path{6f8547c} & E11-C \\
SMoRFFI S1 implemented as a random packet-level split with the lookup-control device decoder, where the preregistration said cross-fitted bound & A & \path{da47fdc} & S1 (implementation committed before S1 was run) \\
SMoRFFI 123-device sensitivity arm & A & \path{2b77364} & E11-C and its sensitivity arm \\
SMoRFFI claim based on the calibrated, not the uncalibrated, bound (post-hoc estimator correction) & \textbf{H} & \path{ea56d97} & decided after the uncalibrated bound read +0.514 bits on 122 devices but -0.097 bits on 123; the calibrated bound read +3.338 and +3.384 \\
WIDEFT role and hierarchy judgement from metadata & A & \path{45997f0} & ingest gates \\
WIDEFT Bluetooth gate redesigns (channelised detection, noise-debiased envelope statistic, hop segmentation) and the one-shot stop rule & A & \path{2848a14}, \path{14af8b9}, \path{e116171} & Bluetooth exclusion (decided on synthetic data after label-free gate failures) \\
WIDEFT Wi-Fi gate, including a corrected DSSS diagnostic & A & \path{513d10c}, \path{787178f} & Wi-Fi exclusion \\
LoRa corpus choice, roles, estimands and verdict sentence & P & \path{9a8f416} & LoRa E11 \\
LoRa dechirp front-end freeze, including the rerun at 400 packets per device & A & \path{bdebd06} & LoRa E11 (frozen before the corpus was read) \\
Family bootstrap replaced by leave-one-family-out jackknife & A & \path{ba63e17} & LoRa decomposition (descriptive) \\
LoRa temporal-contiguity surrogate & \textbf{H} & \path{7cfd37c} & designed after \path{dc67dc3} reported the LoRa result \\
LoPy4 sensitivity arms and the direction-reversal rule & A & \path{3394357} & LoPy4 arms \\
Freeze on further experiments & A & \path{3394357} & all remaining work is audit only \\
\end{longtable}
}
}

\subsection{Status of every analysis}\label{status-of-every-analysis}

The table lists every analysis the paper reports, its outcome, when its design was fixed relative
to its result, and the role it is allowed to play. Outcome labels follow the project ledgers:
\emph{supported} means the stated comparison held under the stated protocol, \emph{proven} is reserved for
synthetic checks with a known ground truth, \emph{inconclusive} means the data cannot decide,
\emph{refuted} means the check failed, and \emph{no signal} means an ingest gate stopped the analysis before
any identity estimate.

{\def\LTcaptype{none} 
{\small
\begin{longtable}[]{@{}
  >{\raggedright\arraybackslash}p{(\linewidth - 6\tabcolsep) * \real{0.2500}}
  >{\raggedright\arraybackslash}p{(\linewidth - 6\tabcolsep) * \real{0.2500}}
  >{\raggedright\arraybackslash}p{(\linewidth - 6\tabcolsep) * \real{0.2500}}
  >{\raggedright\arraybackslash}p{(\linewidth - 6\tabcolsep) * \real{0.2500}}@{}}
\toprule\noalign{}
\begin{minipage}[b]{\linewidth}\raggedright
Analysis
\end{minipage} & \begin{minipage}[b]{\linewidth}\raggedright
Outcome
\end{minipage} & \begin{minipage}[b]{\linewidth}\raggedright
Design fixed
\end{minipage} & \begin{minipage}[b]{\linewidth}\raggedright
Role in the paper
\end{minipage} \\
\midrule\noalign{}
\endhead
\bottomrule\noalign{}
\endlastfoot
Synthetic ceiling and decoder saturation & proven (synthetic) & not preregistered; ground truth known by construction & shows the bound can approach a known ceiling \\
Synthetic permutation null and chain-rule residual & proven (synthetic) & not preregistered; ground truth known by construction & protocol check for the decoder \\
Synthetic cross-PHY matched-null comparison & supported (synthetic only) & before the real replication, not preregistered & motivates, does not replace, a real replication \\
Synthetic cross-PHY decomposition consistency & refuted & before the real replication, not preregistered & restricts decomposition claims to one front-end per PHY \\
WiSig generation grouping vs matched random groupings & supported (comparative) & revised with WiSig in view, then frozen & discovery \\
WiSig non-equalized preambles & supported (direction kept) & amended before the run & sensitivity \\
WiSig monotonic generation trend & inconclusive, withdrawn & after an earlier result & negative result \\
Receiver-chain tap-to-tap comparison & not licensed by the audit & audit rule frozen before a prospective test & restricted result \\
LoRa model grouping vs matched random groupings & supported (directional replication) & preregistered & replication \\
LoRa LoPy4 model-label and chipset-label arms & no reversal & reversal rule fixed before either arm was read & sensitivity only \\
LoRa temporal-contiguity surrogate & did not reproduce the effect & post hoc & limitation check \\
SMoRFFI device identity, 122 devices & supported, with capture-session qualifier & measurement preregistered; calibrated basis chosen post hoc & same-model control \\
SMoRFFI address-block grouping & no detectable excess & preregistered & same-model control \\
SMoRFFI 123-device arm & calibrated bound and grouping result unchanged in substance & amended before the run & appendix sensitivity \\
WIDEFT Bluetooth subset & no signal, corpus closed & ingest gate and stop rule fixed before the one-shot run & exclusion \\
WIDEFT Wi-Fi subset & no signal, corpus closed & ingest gate fixed before the run & exclusion \\
\end{longtable}
}
}

\subsection{Numbers quoted in the abstract}\label{numbers-quoted-in-the-abstract}

Every number in the abstract comes from one of four roles: the frozen audit estimator (16-dimensional
tap-invariant features with a multilayer perceptron, used for every hierarchy comparison), a
protocol check that uses the same decoder on fixed physical features, a high-capacity decoder used
only to show that the synthetic ceiling is reachable, or a sensitivity arm that is never used for a
claim.

{\def\LTcaptype{none} 
{\small
\begin{longtable}[]{@{}
  >{\raggedright\arraybackslash}p{(\linewidth - 6\tabcolsep) * \real{0.2500}}
  >{\raggedright\arraybackslash}p{(\linewidth - 6\tabcolsep) * \real{0.2500}}
  >{\raggedright\arraybackslash}p{(\linewidth - 6\tabcolsep) * \real{0.2500}}
  >{\raggedright\arraybackslash}p{(\linewidth - 6\tabcolsep) * \real{0.2500}}@{}}
\toprule\noalign{}
\begin{minipage}[b]{\linewidth}\raggedright
Number
\end{minipage} & \begin{minipage}[b]{\linewidth}\raggedright
Quantity
\end{minipage} & \begin{minipage}[b]{\linewidth}\raggedright
Estimator and role
\end{minipage} & \begin{minipage}[b]{\linewidth}\raggedright
Source
\end{minipage} \\
\midrule\noalign{}
\endhead
\bottomrule\noalign{}
\endlastfoot
5.910 bits (high-capacity CNN), 98.5\% & held-out bound on the synthetic 6-bit ceiling & high-capacity convolutional decoder; ceiling check only & \path{results/a17_saturation_4x.json} \\
+0.072 bits & mean chain-rule residual, five seeds, synthetic & fixed 12-dimensional physical features with the same MLP decoder; protocol check, not the audit front-end & \path{results/e1_scaling.json} \\
+0.060 bits & chain-rule residual on WiSig & 16-dimensional tap-invariant features with MLP, the audit front-end; validation & \path{results/e7_wisig_full.json} \\
all 999, p = 0.001 (WiSig) & real generation grouping vs matched random groupings & frozen audit estimator; primary & \path{results/e11_final_999.json} \\
all 999, p = 0.001 (LoRa) & real model grouping vs matched random groupings & frozen dechirp front-end with the same decoder protocol; preregistered primary & \path{results/v2_lora_e11.json} \\
+0.296 bits & LoRa decomposition residual & frozen dechirp front-end; descriptive, limits interpretation & \path{results/v2_lora_e11.json} \\
all 999 (two arms) & LoPy4 model-label and chipset-label arms & frozen dechirp front-end; sensitivity only & \path{results/v2_lora_e11_lopy4.json}, \path{results/v2_lora_e11_lopy4_chipset.json} \\
+3.338 bits, 6.931 bits, 122 & SMoRFFI temperature-calibrated device bound and label entropy & 16-dimensional tap-invariant features decoded with the larger \(mlp_{\text{big}}\) device decoder (the same decoder as the device-posterior lookup control) on a random 55/15/30 packet split, with temperature fitted on the validation split; this is not the cross-fitted audit estimator & \path{results/e11c_smorffi_control.json} \\
\end{longtable}
}
}

\section{Appendix B. Limitations: WIDEFT Bluetooth Exclusion}\label{appendix-b.-limitations-wideft-bluetooth-exclusion}

\textbf{WIDEFT Bluetooth was excluded before identity analysis.} A preregistered synthetic calibration, run under
WIDEFT's capture conditions (56 MS/s wideband recording, hop-segmented channelisation), established 11 dB as the
lowest in-channel SNR at which constant-envelope (GFSK) and EDR-like bursts were reliably discriminated. Among
400 bursts drawn at random with a fixed seed from the 8,100 Bluetooth bursts, 200 (50.0\%) contained no hop
segment above this threshold, exceeding the preregistered 20\% corpus-failure limit. No device- or family-level
identity score was computed, and the distribution of failures across devices or families was not inspected,
before exclusion. Because post-hoc SNR filtering could alter device and family composition (captures were taken
per device with hand-set antenna orientation), no partial-corpus analysis was performed. The exclusion therefore
reflects an incompatibility between this corpus and the preregistered front-end, not evidence about the presence
or absence of hierarchical identity structure in WIDEFT. The gate required two label-free redesigns before this
outcome (wideband energy detection failed on hop-channel bursts; raw envelope CV misclassified GFSK at 13 dB);
both redesigns were decided on synthetic data and committed before being applied.

\end{document}